\documentclass[11pt]{article} 
\usepackage{amsmath,amsfonts,graphicx,mathrsfs} 

\newcommand{\be}{\begin{align}}
\newcommand{\ee}{\end{align}}
\newcommand{\bear}{\begin{eqnarray}}
\newcommand{\eear}{\end{eqnarray}}

\newcommand{\sdet}{\mathrm {Sdet}}

\newcommand{\ba}{\begin{array}}
\newcommand{\ea}{\end{array}}

\newcommand{\nn}{\nonumber}
\newcommand{\diag}{\textrm{diag}}

\newcommand{\zbar}{\bar{z}}

\begin{document}
\title{\bf On the mean density of complex eigenvalues for an ensemble of
random matrices with prescribed singular values}
\author{Yi Wei $^{1,2}$ and Yan V. Fyodorov $^{2,1}$\\\\
 $^1$ Isaac Newton Institute for Mathematical Sciences, Cambridge,
UK
\\$^2$ School of Mathematical Sciences, University of Nottingham,
UK}
\date{}

\maketitle \vspace*{-0mm} \abstract{Given any fixed $N \times N$
positive semi-definite diagonal matrix $G\ge 0$ we derive the
explicit formula for the density of complex eigenvalues for random
matrices $A$ of the form $A=U\sqrt{G}$} where the random unitary
matrices $U$ are distributed on the group $\mathrm{U(N)}$
according to the Haar measure.

\section{Introduction}
The question of characterizing the locus of complex eigenvalues
for an $N\times N$ matrix $A$ with prescribed singular values, that is
 eigenvalues $g_i\ge 0,\,\, i=1,\ldots, N$ of $ A^{*} A$,
, with $A^*$ being Hermitian conjugate of $A$, 
  was considered in classical papers by Horn
and Weyl \cite{sing}. In the present paper we provide a kind of
statistical answer to that question. Define
$G=\diag(g_1,\ldots,g_N)\ge 0$, multiply the matrix $\sqrt{G}$ by a
general unitary transformation $U$ from the left and average over
the unitary group $\mathrm{U}(N)$ with the invariant (Haar) measure. This
construction induces a natural measure on the set of matrices
$ A=U\sqrt {G}$ with given singular values $g_i$ and in this way
provides us with the corresponding random matrix ensemble. The
complex eigenvalues $z_i$ of such matrices $A$ will cover
generically an annular domain $R_{min}<|z|<R_{max}$ 
in the complex plane with some density $\rho(z)$. In such 
an approach the statistical
characterization of the locus of complex eigenvalues amounts to
knowing the profile of the ensemble averaged value of that density  
for a given set of singular values.  It is easy to understand that such mean density
 can depend only on $|z|$. The corresponding explicit formula is
provided in {\bf Theorem 2.1} which is the main result of
the paper.

Apart from being a rather nontrivial mathematical problem,
understanding the statistical properties of complex eigenvalues of
the above-mentioned ensemble is motivated by its applications in the
domain of Quantum Chaotic Scattering. In this capacity the problem
attracted attention for some time, and a few partial results were
obtained previously in several limiting
cases\cite{trunc,prel,FSJETP,rev03,yb05}. Below we give a brief
description of the physical context related to the problem.

 One of  very useful instruments in the
analysis of classical Hamiltonian systems with chaotic dynamics
are the so-called area-preserving chaotic maps, see e.g.
\cite{Haakebook} and references therein. They appear naturally either as a mapping of
the Poincar\'{e} section onto itself, or as a result of a
"stroboscopic" description of Hamiltonians which are periodic
functions of time. Their quantum mechanical analogues are unitary
operators which act on Hilbert spaces of finite large dimension
$N$, and are often referred to as evolution, scattering or Floquet
operators, depending on the given physical context. Their
eigenvalues consist of $N$ points on the unit circle
(eigenphases). Numerical studies of various classically chaotic
systems suggest that the eigenphases conform statistically quite
accurately the results obtained for unitary random matrices of a
particular symmetry (Dyson circular ensembles).

Let us now imagine that a system represented by
 a chaotic map ("inner world") is embedded in a
larger physical system ("outer world") in such a way that it
describes particles which can come inside the region of chaotic
motion and leave it after some time via $M$ open channels. Models
of such type appeared in various disguises for example, in
 \cite{KolRev,JSB,OKG,Prange}
 and most
recently discussed in much detail in relation to properties of dielectric 
microresonators in \cite{KNS}. A natural
mathematical framework allowing to deal efficiently with such a
situation was suggested in \cite{FSJETP}, see also \cite{rev03}
and we mention here only its gross features. For a closed quantum
system characterized by a wavefunction $\Psi$
 the "stroboscopic" (discrete-time) dynamics amounts to a
linear unitary map $\Psi(n)\to\Psi(n+1)$, such that
$\Psi(n+1)=\hat{u}\Psi(n)$. The unitary evolution operator
$\hat{u}$ describes the "closed" inner state domain decoupled both
from input and output spaces. Then a coupling that makes the
system open must convert the evolution operator $u$ to a
contractive operator $\hat{A}$ such that ${\bf
1}-\hat{A}^{*}\hat{A}\ge 0$. It is easy to show that one can
always choose $\hat{A}=\hat{u} \sqrt{{\bf
1}-\hat{\tau}\hat{\tau}^{*}}$ where the matrix $\hat{\tau}$
is a rectangular $N\times M,  M \le N $ diagonal with the entries
$\tau_{ij}=\delta_{ij}\tau_j\, , \, 1\le i\le N\, ,\, {1\le j\le
M} ,\quad 0\le \ \tau_j\le 1$. With $\hat{u}$ replaced by $\hat{A}$,  the equation
$\Psi(n+1)=\hat{A}\Psi(n)$ then describes an irreversible decay of
any initially prepared state $\Psi(0)\ne 0$, assuming that 
 external input is absent during the subsequent evolution.
The complex eigenvalues $z_k$ of the operator $\hat{A}$ all belong
to the interior of the unit circle $|z|<1$ and play the role of
resonances for the discrete time systems. Let us mention that
 various aspects of
 resonances associated with quantum chaotic open
maps recently attracted considerable attention\cite{KNS,Nonnem}.

The relation with the random matrix construction
 employed in this paper is now obvious.
It amounts to replacing the true
evolution operator  
$\hat{u}$ with its random matrix analogue $u$ taken from the Circular Unitary Ensemble (CUE)
 in accordance with the standard
ideas of Quantum Chaos,  and to denoting $G=
\sqrt{{\bf 1}-\hat{\tau}\hat{\tau}^{*}}$.
 In this way the task of studying resonances is reduced
to investigating eigenvalues of  random matrices $ A$ of the specified type.
\footnote{
Let us however note that the case most frequently encountered in
 direct physical applications actually
corresponds to choosing $u$ 
to be {\it unitary symmetric} \cite{KNS} taken from COE.
 Such a choice reflects the inherent 
time-reversal invariance typical for the
the closed quantum chaotic system. A somewhat simpler choice of unrestricted unitary random matrices from CUE
corresponds to system with broken time reversal invariance.}

The first significant random matrix result on such matrices $A$ 
seems to have appeared in
\cite{trunc} where the authors considered so-called "truncations" of  random unitary
matrices. In our notations that case is equivalent to
taking $g_1=g_2=\ldots g_M=0$, with all the rest
$N-M$ of $g_i$ being equal to unity. 
In \cite{FSJETP} the mean density of complex eigenvalues was
derived in the limit $N\to \infty$, with
$M<\infty$ being fixed and all $g_i\le 1$.   The
work \cite{rev03} provided some 
general results
on the joint probability
density of all $N$ complex eigenvalues $z_i$, as well as a few
formulae for the few-point correlation functions (the so-called marginal
distributions) of the eigenvalue densities. Those formulae however only
lead to
 explicit managable expressions again 
 in the same limiting case as in \cite{FSJETP}.
As to the results
valid for arbitrary finite $N$, only the simplest particular case
$g_1=g_2=\ldots=g_{N-1}=g,\, g_N<g$ was so far addressed by a
variety of methods, see \cite{yb05} for the most recent account
and \cite{prel} for an early consideration.

\section{Statement of the Main Results}

Our results show that the ensemble-averaged eigenvalue density function $\rho(z)=\Psi(|z|^2)$, i.e. 
indeed depends only on $|z|$. To write the function $\Psi$ explicitly we need to introduce
a few notations.

 Let $\mathrm{s}^l$ be the $l$-th
order elementary symmetric polynomials of $g_i,\ i=1,\dots,N$, e.g. $\mathrm{s}^0=1,
\ \mathrm{s}^1=\sum_{i=1}^Ng_i,\ \mathrm{s}^2=\sum_{i<j}^Ng_ig_j,\ \dots$, etc.
Let us denote $\mathrm{s}^l_{[i_1,i_2,\dots]}=\mathrm{s}^l|_{g_{i_1}=g_{i_2}=\dots=0}$.
Define the following functions of $\{g_i\}$ and the complex variable $z$:
\begin{align}
\mathrm{F}_-(g_i)&=-\frac{1}{{{\prod'}_{j=1}^{N}} (g_i-g_j)}N g_i^{N-1}
\sum\limits_{l=0}^{N-1}\mathrm{s}^l_{[i]}|z|^{2(-l-1)}\,\frac{l}{\left(\!\!\!\begin{array}{c}
N\!-\!1 \\ l \end{array}\!\!\!\right)}\\
\mathrm{F}_+(g_i)&=\mathrm{F}_-(g_i)+\mathrm{F}_\Delta(g_i)\\
\label{eq:fdelta}\mathrm{F}_\Delta(g_i)&
=\frac{1}{{\prod\limits_{j=1}^{N}}^\prime (g_i-g_j)}(g_i-|z|^2)^{N-2}
\sum\limits_{l=0}^{N-1}\mathrm{s}^l_{[i]}|z|^{-2(l+1)}\frac{1}{\left(\!\!\!\begin{array}{c}
 N \!-\!1 \\ l \end{array}\!\!\!\right)}
\bigg[lg_i+(N-1-l)|z|^2\bigg]\nn\\
&=\frac{(g_i-|z|^2)^{N-2}}{{\prod\limits_{j=1}^{N}}^\prime (g_i-g_j)}\int_0^\infty
\frac{Ndt}{(1+t)^{N+2}}\det\bigg(1+\frac{t}{|z|^2}G_{[i]}\bigg)
\bigg[N-t+\frac{g_i}{|z|^2}(Nt-1)\bigg]\;,
\end{align}
where we used the binomial function \small{$\left(\!\!\begin{array}{c}
N\\l\end{array}\!\!\right)=\frac{N!}{l!(N-l)!}$}.
In Eq.\eqref{eq:fdelta}, we defined a matrix $G_{[i]}=\diag(g_1,\dots,g_{i-1},
g_{i+1},\dots,g_N)$ and used the dash to denote that the index $j$ can not be equal to $i$
in the product. 

The main statement of the paper is that the density of complex eigenvalues
can be written in terms of the functions defined above. More precisely, we state the following\\

\noindent{\bf{Theorem} 2.1} Let $U\in \mathrm{U}(N)$ be an element of the unitary
group and $G=\diag(g_1,\dots,g_N)$ be a fixed positive diagonal matrix, such that
$0<g_1<\dots<g_N<\infty$. Let $U$ be distributed on $\mathrm{U}(N)$ according to the Haar measure.
Then the mean density $\rho(z)$ of complex eigenvalues of the matrix  $A=U\sqrt{G}$
 is given by
\begin{align}\label{eq:sum}
\rho(z)=\Psi(|z|^2)=\frac{1}{N}\sum_{i=1}^N \mathrm{F}_\sigma(g_i)\;,
\end{align}
where $\sigma=+$ for $|z|^2>g_i$, $\sigma=-$ for $|z|^2<g_i$.\\

\noindent{\bf{Remark}}:
Exploiting  that the function $\mathrm{F}_-$ is totally antisymmetric with respect to all $g_i$'s, we
can rewrite the above expression in the form:\\

\begin{align}\label{eq:Psi}
\Psi(|z|^2)=\left\{ \begin{array}{ll} 0 & |z|^2<g_1<g_2\dots<g_N\\
\!\! \frac{1}{ N}\sum\limits_{i=k+1}^{N}\mathrm{F}_\Delta(g_i)& g_1<\dots<g_k<|z|^2<g_{k+1}<\dots<g_N
\\ 0 & g_1<g_2\dots<g_N<|z|^2 \end{array}\right.
\end{align}

\noindent{\bf{Remark}}: For $N>3$, it is not difficult to show that the eigenvalue density function is 'smooth' at
each $|z|=g_i, 1< i< N$, that is it has a continuous derivative. 
When $N=3$, the density function is only 'continuous' at $g_2$ but not 'smooth'.\\

\noindent In the case of degenerate eigenvalues of the matrix $G$ the density function can be derived
from Theorem 2.1 by taking the corresponding limits as shown below. \\

\noindent{\bf{Corollary} 2.2}: Suppose the diagonal matrix $G$ has the following
degeneracies,
\begin{align}\label{eq:degenerate}
g_{k_1}=\dots=g_{k_1+i_1},\ \ g_{k_2}=\dots=g_{k_2+i_2}, \ \ \dots, \ \ g_{k_s}=\dots=g_{k_s+i_s}\;,
\end{align}
which is denoted by the short-hand notation
\begin{align}
{G}=\diag(\dots,[g_{k_1},\dots,g_{k_1+i_1}],\dots,[g_{k_s},\dots,g_{k_s+i_s}],\dots)\;.
\end{align}
Define the following two functions

\begin{align}
\mathrm{f}^{[k,i]}_n(g)\overset{\mbox{\tiny def}}{=}\frac{(g-|z|^2)^{N-2}}
{\prod\limits_{j=1}^{k-1}\prod\limits_{j=k+i+1}^{N}(g-g_j)}\sum\limits_{l=n}^{N-1}
\mathrm{s}^{l-n}_{[k,\dots,k+n]}|z|^{-2(l+1)}\frac{1}{\left(\begin{array}{c}\!\!\!
N\!-\!1\\ l \end{array}\!\!\!\right)}\bigg[lg+(N-1-l)|z|^2\bigg]\;,
\end{align}

\begin{align}
\mathrm{F}^{[k,i]}_\Delta\stackrel{\mbox{\tiny def}}{=}\sum\limits_{n=0}^{i}\frac{(-)^n}{(i-n)!}
\frac{d^{i-n}}{dg^{i-n}_{k+n}}\mathrm{f}^{[k,i]}_n(g_{k+n})\;.
\end{align}
The density function is then given by replacing each $\sum_{n=0}^i\mathrm{F}_\Delta(g_{k+n})$ in
the Theorem by $\mathrm{F}_\Delta^{[k,i]}$ and making substitutions Eq.{\eqref{eq:degenerate}}.\\

\noindent{\bf{Proof}:}
Consider the following sum: $\psi=\sum_{n=0}^i\mathrm{F}_\Delta(g_{k+n})$ when $g_{k}=\dots=g_{k+i}$.
Taking the limit $g_{k+i-1}\to g_{k+i}=g$ yields
\begin{align}
&\lim_{g_{k+i-1}\to g_{k+i}=g}\psi\nn\\
=&\left(\sum\limits_{n=0}^{i-2}
\mathrm{F}_\Delta(g_{k+n})+\frac{d}{dg_{k+i-1}}(g_{k+i-1}-g_{k+i})
(\mathrm{F}_\Delta(g_{k+i-1})+\mathrm{F}_\Delta(g_{k+i}))\right)_{g_{k+i-1}=g_{k+i}=g}\nn\\
=&\left(\sum\limits_{n=0}^{i-2}
\mathrm{F}_\Delta(g_{k+n})+\frac{d}{dg_{k+i-1}}
\left[\mathrm{f}_0^{[k+i-1,1]}(g_{k+i-1})-\mathrm{f}_1^{[K+i-1,1]}
(g_{k+i})\right]\right)_{g_{k+i-1}=g_{k+i}=g}\nn\\
=&\ \mathrm{F}_\Delta^{[k+i-1,1]}|_{g_{k+i-1}=g_{k+i}=g}\;.
\end{align}
In the second step we have exploited the linearity of $\mathrm{s}^l$ as a function of $g$'s.
Next, we take the limit $g_{k+i-2}\to g$. Repeating the procedure $i$ times, we arrive at
\begin{align}\label{eq:single-deg}
\psi|_{g_{k}=\dots=g_{k+i}=g}=\mathrm{F}_\Delta^{[k,i]}|_{g_{k}=\dots=g_{k+i}=g}\;.
\end{align}
Applying the results Eq.{\eqref{eq:single-deg}} to other degeneracies, we obtain the eigenvalue density
function for $U\sqrt{G}$ under the condition Eq.{\eqref{eq:degenerate}}.\hspace*{10mm}$\Box$\\

\noindent{{Example 1}\bf (rank-one deviation from unitary matrix):}\\ In the special case of
$G=\diag(g_1,[g_2,\dots,g_N])$, with $g_2=\dots=g_N=g$, the above procedure leads to
an especially simple formula for the mean eigenvalue density:
\begin{align}\label{eq:rank1}
\Psi=\frac{(|z|^2-g_1)^{N-2}}{(g-g_1)^{N-1}|z|^{2N}}\!\left((N-1)(|z|^{2N}+g^{N-1}g_1)+\!\!
\sum\limits_{k=0}^{N-2}\left[(N-2-k)g+kg_1\right]g^k|z|^{2(N-1-k)}\right)\;.
\end{align}
which coincides with the known result \cite{prel,yb05}.\\

\noindent{\bf{Remark}}: In Eq.{\eqref{eq:rank1}}, as $g\to g_1$, the density function $\Psi\to\infty$
on $[g_1,g]$ and is zero otherwise. On the other hand, the integration of $\Psi$ over the
region $[g_1,g]$ yields one.  We conclude that in this case the density
function is simply $\delta(g-|z|^2)$, as it must be for a random matrix $A=U\sqrt{g}$ which is 
simply proportional to CUE matrix.\\

\noindent {\bf{Remark}}: In fact, in our derivation of the main theorem, we can extend the
domain of $g_i$'s to include the origin, i.e. our formula holds for $g_i\ge0$. We illustrate
this observation in the following example.\\

\noindent{{Example 2}\bf (truncated unitary matrix):}\\
Consider the case  $G=\diag([g_1,\dots,g_M],[g_{M+1},\dots,g_N])$,
where $g_1=\dots=g_M=0$ and $g_{M+1}=\dots=g_N=1$. By Corollary, we can write
the density function of eigenvalues of $A=U\sqrt{G}$ as
\begin{align}\label{eq:truncate}
\Psi=&\ \mathrm{F}^{[M+1,N-M-1]}_\Delta|_{g_1=\dots=g_M=0,g_{M+1}=\dots=g_N=1}\nn\\
\propto&\ (1-|z|^2)^{M-1}\left(\frac{d}{d|z|^2}\right)^M\frac{1-|z|^{2N}}{1-|z|^2}\;.
\end{align}
In the first step, we have extended domains of $g$ to $[0,\infty)$ in Theorem 2.1, and
correspondingly, Corollary 2.2. In fact, when $G=\diag(0\cdot I_M,I_{N-M})$, eigenvalues
of the matrix $A$ in Example 2. coincide with those of $(N-M)\times(N-M) $ lower right sub-block
of a random unitary matrix, also known as the 'truncated'
unitary matrix. Same results as
Eq.{\eqref{eq:truncate}} has been obtained with a  completely different method in \cite{trunc}. \\

\noindent {\bf{Remark}}: As we can obviously always absorb the $\mathrm{U}^N(1)$ phase of
$N\times N$ complex diagonal matrix into $U$, the domain for the matrix $ G$ can
be defined on $\mathbb{C}_1^N$. The eigenvalue density function of $U\sqrt{G}$,
averaged over CUE is then obtained by substituting $g_i\to |g_i|$ into Theorem 2.1.\\

\begin{figure}
\centering
\includegraphics[width=7.5cm, height=6cm]{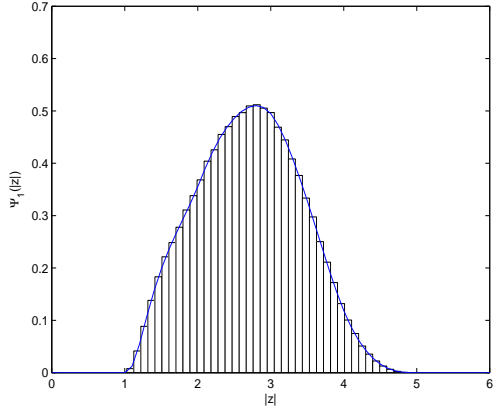}
\caption{Histogram of the radial part of the eigenvalue distribution of $5\times5$
matrices $A=U\sqrt{G}$. Here $G=\diag(1,4,9,16,25)$ and  $U$ is the  Haar-distributed unitary
matrix, with sample size 100000 and bin 0.10. Solid
line represents the function $\Psi_1(|z|)$ as derived from Eq.\eqref{eq:Psi}.}
\label{histo}
\end{figure}

\noindent Finally, we compare our formula Eq.\eqref{eq:Psi} with numerical simulations. To this
end, we choose a fixed diagonal matrix $ G$ and generate unitary matrices according to the Haar measure.
We draw a histogram of the radial part of eigenvalues, $|z|$, of the matrix $U\sqrt{G}$,
see Fig.\ref{histo} To compare with the histogram, we define the appropriately modified density function
$\Psi_1(|z|)=2|z|\Psi(|z|^2)$, which is shown by the solid line. From Fig. \ref{histo}, we observe a
 very
good match between our formula Eq.\eqref{eq:Psi} and the results of numerical simulations.

\hspace{75mm}

\section{Main steps of the proof}
\subsection{Colour-flavour transformation}
Our starting expression is the following formula \cite{FKS,rev96}
for the averaged density of complex eigenvalues for a general
finite-size non-Hermitian random matrix $A$
\begin{align}\label{eq:formula}
\rho(z)=-\frac{1}{\pi}
\lim_{\kappa\to0}\frac{\partial}{\partial{\bar z}}\lim_{z_b\to z}
\frac{\partial}{\partial{z_b}}\Bigg<\frac{\det\left(
\begin{array}{cc}\kappa&i(z-A)\\ i(\bar{z}-A^\dagger)&\kappa\end{array}\right)}{\det\left(
\begin{array}{cc}\kappa&i(z_b-A)\\ i(\bar{z}_b-A^\dagger)&\kappa\end{array}\right)}\Bigg>_U\;.
\end{align}
In our case $A=U\sqrt{G}$, where $U\in \mathrm{U}(N)$, $G=\diag(g_1,\dots,g_N)>0$.
Averaging over the Haar measure on the unitary group $\mathrm{U}(N)$ is denoted by $<\cdots>_U$.

Introduce vectors $S_a=(s_a^i)$ with complex components and their counterparts $\chi_a=(\chi_a^i)$ 
with anti-commuting components (Grassmann variables), for all $i=1,\dots,N$ and $a=1,2$. 
This defines two sets of (graded) vectors
$\psi^i_{a}=\left(\begin{array}{c} s^i_{a}\\ \chi^i_{a}\end{array}\right)$ with $a=1,2$.
The determinants can be represented as integrals over complex variables $s$ and Grassmann variables $\chi$
in the standard way:
\begin{align}\label{eq:int-before-cft}
<\cdots>_U\propto&\int dU \int dS_1 dS_2 \exp\left[-\kappa(S_1^\dagger S_1+S_2^\dagger S_2)
-i(z_bS^\dagger_1S_2+\bar{z}_b S^\dagger_2S_1)\right]\nn\\
&\int d\chi_{1}d\bar{\chi}_1d\chi_{2}d\bar{\chi}_2\
\exp\left[-\kappa(\chi_1^\dagger \chi_1+\chi_2^\dagger \chi_2)
-i(z\chi^\dagger_1\chi_2+\bar{z}\chi^\dagger_2\chi_1)\right]\nn\\
&\ \exp i\left[\bar\psi_{1}^iA_{ij}\psi_2^j+\bar\psi_{2}^iA_{ij}^\dagger\psi_1^j\right]\;,
\end{align}
where we defined $dS_1\,dS_2=\prod_{a=1}^2\prod_{i=1}^N d\bar{s}_a^ids_a^i$.
Note that by Eq.{\eqref{eq:formula}}, averaging over the unitary group
in the above expression should be carried out after performing the integral over the graded vectors $\{\psi,\bar\psi\}$.\\

Next we change the order of integration over $\{\psi,\bar\psi\}$ and $\mathrm{U}(N)$, which is possible due
to the fact that $\mathrm{U}(N)$ is compact and the integral is bounded. The integration over the unitary
group can be  performed explicitly by exploiting the colour-flavour transformation discovered 
by Zirnbauer \cite{cft}:
\begin{align}\label{eq:cft}
\int dU \exp \mathrm{i}\left[\bar\psi_{1}^iA_{ij}\psi_2^j+\bar\psi_{2}^iA_{ij}^\dagger\psi_1^j\right]
=\int D(Q,\tilde{Q})\ \exp \mathrm{i}
\left[ \bar{\psi}_{1}^iQ\psi_{1}^i+g_i\bar{\psi}_{2}^i\tilde{Q}\psi_{2}^i\right]\;.
\end{align}
Such a transformation trades the integration over $\mathrm{U}(N)$, where N can be an arbitrary
large integer for the integration over a considerably simpler $2\times2$ graded matrices $Q$ defined as
\begin{align}\label{eq:cfta}
Q=\left(\begin{array}{cc} q_b&\eta_1\\ \eta_2&q_f\end{array}\right),
\hspace*{5mm} \tilde{Q}=\left(\begin{array}{cc} \bar{q}_b&\sigma_1\\ \sigma_2&-\bar{q}_f\end{array}\right)\;.
\end{align}
Such $Q$ belongs to a Riemannian symmetric superspace \cite{sym-space} of the type $\mathrm{AIII|AIII}$.
Here, $\eta$'s and $\sigma$'s are anti-commuting Grassmann variables. The so-called boson-boson
and fermion-fermion blocks of $Q$ are given by
\begin{align}\label{eq:cftb}
q_b\in\mathrm{U(1,1)/U(1)\times U(1)}=\mathrm{H}^2\ {\mathrm{and}}\ q_f\in\mathrm{U(2)/U(1)\times U(1)=S^2}\;.
\end{align}
The invariant measure on this domain is defined as
\begin{align}
D(Q,\tilde Q)=\sdet^N(1-\tilde QQ)dQd\tilde{Q}\;.
\end{align}
where $dQd\tilde{Q}=d^2q_bd^2q_fd\sigma_1d\sigma_2d\rho_1d\rho_2$.

After the colour-flavour transformation Eq.{\eqref{eq:cft}}, we get
\begin{align}\label{eq:qcft}
<\cdots>_U\ \propto F(\kappa)=\int d\bar{\psi}d\psi\int D(Q,\tilde Q)\exp-(\bar{\psi}_1^i,\bar{\psi}_2^i)
\left(\begin{array}{cc}\kappa-iQ&\mathrm{i}Z \\ & \\ \mathrm{i}\bar{Z} &
\kappa-ig_i\tilde{Q}\end{array} \right)\left(\begin{array}{l}\psi_1^i\\ \psi_2^i\end{array}\right)\;,
\end{align}
where $d\bar{\psi}d\psi=dS_1^2dS_2^2d\chi_{1}d\bar{\chi}_1d\chi_{2}d\bar{\chi}_2$ and we defined
\begin{align}
Z=\left(\begin{array}{cc}z_b&0\\0&z\end{array}\right)\ \ \mathrm{and}\ \
\bar{Z}=\left(\begin{array}{cc}\zbar_b&0\\0&\zbar\end{array}\right)\;.
\end{align}
For a fixed complex number $z$ and a given diagonal matrix $ G$, the integral in Eq.{\eqref{eq:qcft}},
 defines a function $F(\kappa)$ of the variable $\kappa$. Using
the integral representation for Bessel functions we can show that $F(\kappa)$ is analytic in
the half plane $\mathrm{Re}\ \kappa\!>\!0$.

\subsection{Integration over $ Q$ and analytic continuation}

Direct evaluation of the integral over
the graded matrix $Q$ followed by the integration over $\psi$ in Eq.{\eqref{eq:qcft}} is very
difficult. In fact, it is already a highly involved task in a much simpler case where $Q$ is a complex
number and $\psi$ is a complex vector, see \cite{yb05} for the 
corresponding calculation in such a case. A natural way out could be changing the
order of integration in Eq.{\eqref{eq:qcft}} in order to integrate first over $\{\psi,\bar\psi\}$
by using the standard Gaussian integral formula for graded vectors. However, extra care must be
taken in performing such a change. To understand this consider the integral involving the boson-boson 
part of the supermatrix
$Q$ and the complex vectors $S_1$ and $S_2$,
\begin{align}\label{eq:ibosonic}
I_{\mathrm{bosonic}}=\int dS_1dS_2\int_{|q_b|\le1} dq_b^2\
\mathrm{e}^{-\kappa(S_1^\dagger S_1+S_2^\dagger S_2)-(iz_bS_1^\dagger S_2+i\bar{z}_bS_2^\dagger S_1)
+iq_bS_1^\dagger S_1+i\bar{q}_bg_i\bar{S}_2^iS_2^i}\;,
\end{align}
where we have omitted the trivial Grassmann integrals. Changing the order of integration in
Eq.\eqref{eq:qcft} we arrive at
\begin{align}\label{eq:itilde}
\tilde{I}_{\mathrm{bosonic}}=\int_{|q_b|\le1} dq_b^2\int dS_1dS_2\
\mathrm{e}^{-\kappa(S_1^\dagger S_1+S_2^\dagger S_2)-(iz_bS_1^\dagger S_2+i\bar{z}_bS_2^\dagger S_1)
+iq_bS_1^\dagger S_1+i\bar{q}_bg_i\bar{S}_2^iS_2^i}\;.
\end{align}
It is clear that $\tilde{I}_{\mathrm{bosonic}}$ is only well-defined in $\mathrm{Re}\,\kappa\in(1,\infty)$.
For $\kappa\to 0$, which is the limit we have to perform in the very end
of the calculation, the integration over the boson-boson
domain forbids changing integration order in Eq.{\eqref{eq:qcft}}.
Actually, such a problem was first noticed in \cite{yb05}, and solved by modifying
in a non-trivial way the domain of integration Eq.\eqref{eq:cftb} over bosonic variables
 in the colour-flavour transformation. After such a modification one can 
actually carry out the required change of integration order for any $\kappa>0$.
We shall however see that one can work in the standard parametrisation Eq.\eqref{eq:cftb} in the 
allowed region $\mathrm{Re} \kappa\!\!>\!\!1$, and then continue to  $ 0<\mathrm{Re}\,\kappa<1$
exploiting analytic properties of the function $F(\kappa)$.

Let us from now on work in the domain $\mathrm{Re}\ \kappa\!\!>\!\!1$.
Substituting the transformation Eq.{\eqref{eq:cft}} into Eq.{\eqref{eq:int-before-cft}},
changing the order of integrations over graded vectors $\{\psi,\bar\psi\}$ and the graded matrix
$Q$ and integrating out $\{\psi,\bar\psi\}$, we get
\begin{align}\label{eq:int-after-cft}
G(\kappa)=&\int dQd\tilde{Q}\ \sdet^N(1-\tilde{Q}Q)\prod^N_{i=1}\sdet^{-1}
\left[\begin{array}{cc}\kappa-iQ&i \left(\begin{array}{cc}z_b&0\\0&z\end{array}\right)\\
& \\ i\left(\begin{array}{cc}\zbar_b&0\\0&\zbar\end{array}\right) &
\kappa-ig_i\tilde{Q}\end{array} \right]\;.
\end{align}
Performing the integration over the supermatrix $Q$ is still a rather involved
technical problem. We provide below
a few comments related to it.

 The Grassmann variables can be integrated out at any stage, and
and we find it convenient to carry out that integration at the very 
beginning. On the other hand, it turns out to be important that the integration over the boson-boson
part of $Q$ (i.e. $q_b$) should be performed before the fermion-fermion part $q_f$. In fact, a quick
inspection of the $q_f$ integrals in Eq.\eqref{eq:int-after-cft} shows that they diverge logarithmically.
However, those logarithmic divergences are actually a
spurious feature of the colour-flavour transformation. The correct
way of treating such divergencies when
 performing any supersymmetric colour-flavour transformation
 is to integrate first over the boson-boson sub-manifold. 
Then, combining all terms which are logarithmically divergent, one can show that
the divergent parts cancel each other and the result is actually finite.

To perform the integration over the boson-boson part of $Q$, we introduce polar coordinates
$q_b=\sqrt{r}e^{\mathrm{i}\theta}$, where $r\in[0,1], \theta\in[0,2\pi]$ so that
$dq_bd\bar{q}_b=drd\theta$. In Eq.\eqref{eq:int-after-cft}, $G(\kappa)$ is defined for
$\mathrm{Re}\ \kappa\!\!>\!\!1$. For simplicity, we focus on the real 
$\kappa>1$. And further more, we assume all $g_i<1$, for $i=1,\dots,N$. 
This is done for convenience only and does not reduce generality 
as we can always scale the $ G$-matrix by the magnitude of the largest eigenvalue,
 and at the end of the calculation to scale it
back. Under these assumptions we can integrate 
over the angular variable $\theta$ by residue theorem. In this way 
the result of the original integration
naturally splits into a sum of contributions from different
residues. 

It is crucial that after integrating over $\theta$, we can show $G(\kappa)$ is analytic
in the half plane $\mathrm{Re}\ \kappa\!\!>\!\!0$. Therefore we are allowed to
make the required
analytic continuation
on $G(\kappa)$ to $\mathrm{Re}\ \kappa\!\!>\!\!0$. Since both $F(\kappa)$ and $G(\kappa)$
are now analytic functions on $\mathrm{Re}\ \kappa\!\!>\!\!0$ and $F(\kappa)=G(\kappa)$ on
$\kappa\!\!>\!\!1$, we conclude that $F(\kappa)=G(\kappa)$. We emphasis that this
continuation is possible to
carry out only after angular integration is performed
under the assumption $\kappa\!>\!1$.

 Knowing that we should take the limit $\kappa\to0$ in the end of the calculation, we can use the condition 
$\kappa<<1$ to simplify significantly the integration over the radial
part of $q_b$. It turns out that one has to distinguish
two essentially different cases: $g_i>|z|^2$ and
$g_i<|z|^2$. Each of these two cases yields different result when integrating
over $q_b$ which
explains why we have to
distinguish $F_+(g_i)$ from $F_-(g_i)$ in the final expression.
Integration over the fermion-fermion part of $Q$ uses certain properties of
elementary symmetric functions but is otherwise straightforward. Finally, taking
derivatives with
respect to
$z_b$ and $z$ and letting
$\kappa\to 0$, we arrive after straightforward but still cumbersome calculations to
 the formula Eq.\eqref{eq:sum}.\\

\section{Open problems}
In conclusion, we would like to mention a few open  problems and possible extensions along 
the lines of the present work. An interesting problem would be to investigate 
the density of complex eigenvalues in the limit 
$N\to \infty$ assuming that the matrix $g$ has a finite limiting density $\nu(g)=\frac{1}{N}\sum_i\delta(g-g_i)$
 of eigenvalues $g_i$ in an interval of the  $g-$axis. A special variant of the problem is to assume that 
$g_i$ are eigenvalues of some random Hermitian matrix with rotationally invariant measure. This case is in fact
equivalent to the so-called Feinberg-Zee problem, see \cite{FZ}, which attracted a considerable interest recently.
We hope to address it in our future publications. 

Another important extension would be to replace matrices $U$ by unitary symmetric random matrices, or 
to take them from some other groups (e.g. orthogonal). The corresponding colour-flavour transformations are known,
but the calculations seem to be extremely challenging technically.

\section*{Acknowledgements}
This research was completed during the 2008 programme "Anderson
Localisation: 50 years after" at the Isaac Newton Institute for
Mathematical Sciences where both authors were supported by
visiting Fellowships. We gratefully acknowledge Prof. B.
Khoruzhenko for many clarifying discussions.
We also are grateful to S. Nonnenmacher and J. Keating for their stimulating 
interest in this work. 
The research in Nottingham was performed in the framework of EPSRC
grant EP/C515056/1"Random Matrices and Polynomials: a tool to
understand complexity".

\end{document}